\documentclass[acmsmall]{acmart}
\AtBeginDocument{%
  \providecommand\BibTeX{{%
    \normalfont B\kern-0.5em{\scshape i\kern-0.25em b}\kern-0.8em\TeX}}}

\usepackage{natbib}
\usepackage{graphicx}

\usepackage{url}
\usepackage{xcolor}

\begin{document}


\title{Trustworthiness Evaluations of Search Results: The Impact of Rank and Misinformation}




\author{Sterling Williams-Ceci}
\author{Michael Macy}
\author{Mor Naaman}




\begin{abstract}
Users rely on search engines for information in critical contexts, such as public health emergencies. Understanding how users evaluate the trustworthiness of search results is therefore essential. Research has identified rank and the presence of misinformation as factors impacting perceptions and click behavior in search. Here, we elaborate on these findings by measuring the effects of rank and misinformation, as well as warning banners, on the perceived trustworthiness of individual results in search. We conducted three online experiments (N=3196) using Covid-19-related queries to address this question. We show that although higher-ranked results are \textit{clicked} more often, they are not more \textit{trusted}. We also show that misinformation did not change trust in accurate results below it. However, a warning about unreliable sources backfired, decreasing trust in accurate information but not misinformation. This work addresses concerns about how people evaluate information in search, and illustrates the dangers of generic prevention approaches.

\end{abstract}

\maketitle








\section{Introduction}
People continue to rely on search engines for critical types of information seeking~\cite{Chang2023, Zade2022, Song2022, Metaxa2019, Pogacar2017, Epstein2015}, including for health-related information~\cite{Song2022, Pogacar2017} and disaster response~\cite{palen2011supporting}.
During the recent Covid-19 pandemic, users turned to search engines for information about the disease and its treatment~\cite{Negrone23, Gallotti2020, Jayasinghe2020}. Reliance on search engines for critical information raises increased concerns about the potential threats to public health and wellbeing if users misjudge the trustworthiness of the information they encounter.
Understanding how people evaluate the credibility of information in online search is critical because information from search engines can distort users' knowledge of topics~\cite{Song2022, Pogacar2017, Kay2015} and could potentially change their real-world behavior~\cite{EpsteinACM2017, Epstein2015}. The ways that users evaluate the trustworthiness of information in online search
has been an ongoing interest in the CHI community and beyond (e.g.,~\cite{haussler2023users, Wang2023, Azzopardi2021, Proboyekti2022, Fell2020, Chirag2018, Pan2007}).

One frequently-raised concern is that users interpret search results’ rank as an indicator of the information’s credibility and trustworthiness~\cite{Metaxa2019, Epstein2015}. Rank has been shown to correlate strongly with the probability of being clicked, for searches in general~\cite{Haas2017, Glick2014, Pan2007} and health-related queries in particular~\cite{Pogacar2017, Kammerer2014, Pan2007, Eysenbach2002}. However, these studies have not directly tested whether trust underlies the rank-click relationship: do people believe that higher-ranked information is more trustworthy?

The rank-trust relationship is especially important to understand given the presence of misinformation in high-ranked search results~\cite{Zade2022, Juneja2021, White2014}. 
If the relationship exists, it could mislead search users into perceiving highly ranked misinformation as trustworthy. 
Furthermore, past studies have suggested that exposure to misinformation can cause people to doubt the validity of accurate information shown afterward~\cite{VanDerMeer2023, Song2022, Rapp2018}. This hypothesis is particularly relevant, but not well-explored, in the online search context where accurate results may naturally appear below high-ranked misinformation. 

Another important factor comes into play in the search environment: warning banners. To combat the potential risks associated with misinformation, search engines such as Google show warning banners at the top of result pages when answers to the queries are uncertain~\cite{Ghaffary2021, Sullivan2021}. 
For example, one warning used by Google alerts people to the possibility that the information is “changing quickly.” 
Fact-checking research has shown mixed effects of both general~\cite{VanDerMeer2023, Roozenbeek2022, Clayton2019} and item-specific~\cite{Aslett2022, Clayton2019, Pennycook2020} information reliability warnings in contexts such as news evaluation, but no studies to our knowledge have examined how warning banners affect perceptions of results’ trustworthiness on a search results page. 

To address these research gaps, we performed three online experiments with 3196 participants, measuring users' evaluations of search results in the context of Covid-19. 
Across the various experiments, we randomized the rank of results; manipulated whether misinformation was present; and tested whether inclusion of different warning banners impacted the evaluations. 
The experiments used different Covid-19 queries varying in medical community consensus at the time of the study, attributed the results to two different search engines, and used different types of sources associated with the results to control for source recognition in the evaluations. 
Participants were asked to click one result that they would choose to answer the query, then to rate one to three semi-randomly selected results from the page on a scale of perceived trustworthiness.


Our findings were consistent across all three experiments: we found no relationship between rank and trust evaluations for accurate information. 
Further, we show that misinformation is highly distrusted in comparison to accurate information, even when it is highly ranked. 
Additionally, we show that the presence of high-ranked misinformation does \textit{not} harm the perceived trustworthiness of accurate results below it. 
Interestingly, a warning banner about the unreliability of the results’ sources had a backfire effect, decreasing trustworthiness ratings of accurate results but not misinformation results.
The findings offer clear implications for the design of algorithmic or visual interventions in search environments.

\section{Background and Related Work}

Prior research in multiple disciplines, including human-computer interaction, had touched on related themes to our work here. These efforts examined the impact of rank on search behavior, looked at the cognitive impacts of misinformation in online systems, and studied fact-checking strategies to combat misinformation. 

\subsection{Impact of Rank in Search}
Multiple studies have found that the rank of results in search is correlated with their probability of being clicked~\cite{Pogacar2017, Haas2017, Glick2014, Kammerer2014, Pan2007, Eysenbach2002}. Several explanations have been proposed, many of which relate to Cognitive Miser theory~\cite{Operario2002}. The theory posits that humans avoid using mental resources when making decisions, which leads them to rely on heuristics. 
For example, Azzopardi~\cite{Azzopardi2021} describes clicking on high-ranked results as a “satisficing” behavior by cognitive misers to avoid scrolling through countless results and considering an overwhelming amount of information (p. 32). In support of this hypothesis, Pan et al.~\cite{Pan2007} found that people are more likely to click on high-ranked results even when these results were independently judged as being the least relevant ones in the list.

While the cognitive miser hypothesis is widely accepted, an additional factor may play a role in users' decisions to click high-ranked results: 
users may interpret high rank as an indication that they can \textit{trust} the expertise of the source and the credibility of the content. 
When online search is used to obtain health-related information, the consequences of the information we take away can be very serious~\cite{SwireThompson2020, palen2011supporting}. 
Higher consequences should decrease our tendency to be cognitive misers and lead us to evaluate the information more carefully. Nonetheless, studies have shown that rank is still predictive of click probability for health-related queries \cite{Pogacar2017, Kammerer2014, Pan2007, Eysenbach2002}. 
For these searches in particular, it is plausible that people click high-ranked results in part based on the assumption that rank is indicative of trustworthiness, since finding the right answer has higher stakes for their health and wellbeing than in searches for non-consequential information. 
If this assumption exists among search users, it is concerning: individuals searching for health information may be especially vulnerable to high-ranked misinformation and may discount accurate information from credible sources appearing lower on the page.
Taken together, it is important to understand the impact of rank on trustworthiness evaluations, as we do in this work. 

Despite a long history of studies of trust in search, the link between rank and trustworthiness evaluations has not been studied directly. 
Previous studies have been unable to test whether rank influences the perceived trustworthiness of information in online search for multiple reasons. 
Studies have relied on click-rates as a proxy instead of directly measuring perceptions of trustworthiness~\cite{Haas2017, Kammerer2014}. For example, Pan et al.~\cite{Pan2007} and Schultheis et al.~\cite{Schultheis2018} found that click probability was correlated with the ranking of results more than with the relevance to the query, but they did not measure whether participants also trusted higher-ranked information more than the information in results that were ranked lower. 
As Pan et al.~\cite{Pan2007} titled their paper ``In Google We Trust'', other researchers had cited it as evidence that rank influences perceived trustworthiness of search results. However, Pan et al.'s title instead referenced their finding that people trust Google's ranking, which affected perceived \textit{relevance} of the search results, not the perceived trustworthiness of the information. 
Haas and Unkel~\cite{Haas2017} more directly examined how people evaluate the credibility of individual search results by asking participants to rate individual results on credibility scales -- but they did not manipulate the results' rank or otherwise build on it in their experiment. Their results did show that users' evaluations are indeed sensitive to the information available for individual results, such as whether the source is a reputable website, a factor we include in our study as well. 

Given this gap, we are interested here in formally testing whether rank impacts trustworthiness evaluation as well as click likelihood. 
Since the rank-click relationship has been documented in past research, including for health-related queries, we first propose this baseline hypothesis:

\textit{\textbf{H0:} An increase in the rank
of a search result will be associated with an increase in the probability of it being clicked.}

Past literature has shown the importance of trust when seeking information online in high-stakes contexts~\cite{SwireThompson2020, Pogacar2017, palen2011supporting}. Given that our experiments use queries about Covid-19 vaccine safety, we hypothesize that people will trust results they choose to click more than those they do not click. We note, though, that our experiment setup only allows us to test this relationship in an observational and not causal way; see Section~\ref{sec:limitations} for details.

\textit{\textbf{H1}: Controlling for rank, participants will rate the trustworthiness of results they had clicked higher than ones they had not clicked.}

However, the central question is whether people \textit{trust} high-ranked results more than lower-ranked ones. Our experiments will also allow us to test this hypothesis, which has been assumed in prior work~\cite{Metaxa2019, Epstein2015} yet not directly tested:

\textit{\textbf{H2}: An increase in the rank of a search result will be associated with the result being rated as more trustworthy.}




\subsection{Impact of Misinformation in Trustworthiness Evaluation}

If people infer trustworthiness from the rank of results on a search page, misinformation becomes especially concerning. Search engine audits have found that misinformation appears in results on search pages with alarming frequency~\cite{Zade2022, Juneja2021, White2014} including in high ranks on the search page~\cite{Juneja2021, White2014}, which carries especially significant consequences for public health challenges such as Covid-19~\cite{SwireThompson2020, Hargittai2007}. 
Recent studies have found that exposure to false or misleading search results is associated with an increased likelihood of ill-advised medical decisions~\cite{Song2022, Abualsaud2019, Pogacar2017}. 
Swire-Thompson and Lazer~\cite{SwireThompson2020} concluded that “Misinformation concerning health has particularly severe consequences with regard to people's quality of life and even their risk of mortality” (p. 443).

The problem may go beyond the willingness to believe search results containing misinformation~\cite{Gallotti2020}. While some studies have found that people are good at identifying blatant misinformation~\cite{PennycookAndOthers2018}, researchers have warned that exposure to prominently displayed misinformation may also undermine trust in accurate search results from reliable sources~\cite{VanDerMeer2023, Song2022, Rapp2018}. Numerous experiments have found that misinformation exposure can significantly shift people’s beliefs, opinions, and perceptions of the credibility of information~\cite{PennycookAndOthers2018, Rapp2018}. Specifically, van der Meer et al.~\cite{VanDerMeer2023} found that exposure to misinformation in news stories decreases the perceived credibility of accurate information shown afterward. 
In the context of search, Song and Jiang~\cite{Song2022} found that exposure to a higher number of search results with misinformation was associated with a decreased likelihood of clicking on accurate results in the same list, and (along with Pogacar et al.~\cite{Pogacar2017}) a higher rate of inaccurate medical decisions. 
Taken together, these studies show the potential impact of misinformation in search, and its interactions with accurate information in influencing people's knowledge. 
However, the impact of misinformation's mere presence on trustworthiness perceptions of accurate information at the search result level remains an open question.

We address this gap in prior work by manipulating whether misinformation appears in search result pages in our experiments. 
Specifically, we are interested in whether exposure to a high-ranked piece of misinformation causes people to doubt the veracity of accurate information presented below it. 
We first test whether people recognize when a result contains misinformation. We hypothesize that participants in our main studies will click our misinformation results less often and rate them as less trustworthy than the accurate results, controlling for rank:

\textit{\textbf{H3a}: A result containing misinformation will be less likely to be clicked than a result containing accurate information, controlling for rank.}

\textit{\textbf{H3b}: A result containing misinformation will be trusted less than a result containing accurate information, controlling for rank.}

Assuming participants recognize misinformation in search, does it indeed cause them to doubt the credibility of accurate information presented below? 
Given the evidence above, and given that exposure is guaranteed when misinformation is shown in a high-ranked search result (since people scan the search result page from top to bottom and fixate on the top five results~\cite{Hotchkiss2005}), we hypothesize that: 

\textit{\textbf{H4a}: An accurate result will be less likely to be clicked when the result immediately above it contains misinformation, compared to the control condition in which only accurate results are shown.}

\textit{\textbf{H4b}: An accurate result will be rated as less trustworthy when the result immediately above it contains misinformation, compared to the control condition in which only accurate results are shown.}

\subsection{Impact of Warning Banners in Trustworthiness Evaluation}
\begin{figure}[ht]
  \centering
  \includegraphics[width=120mm]{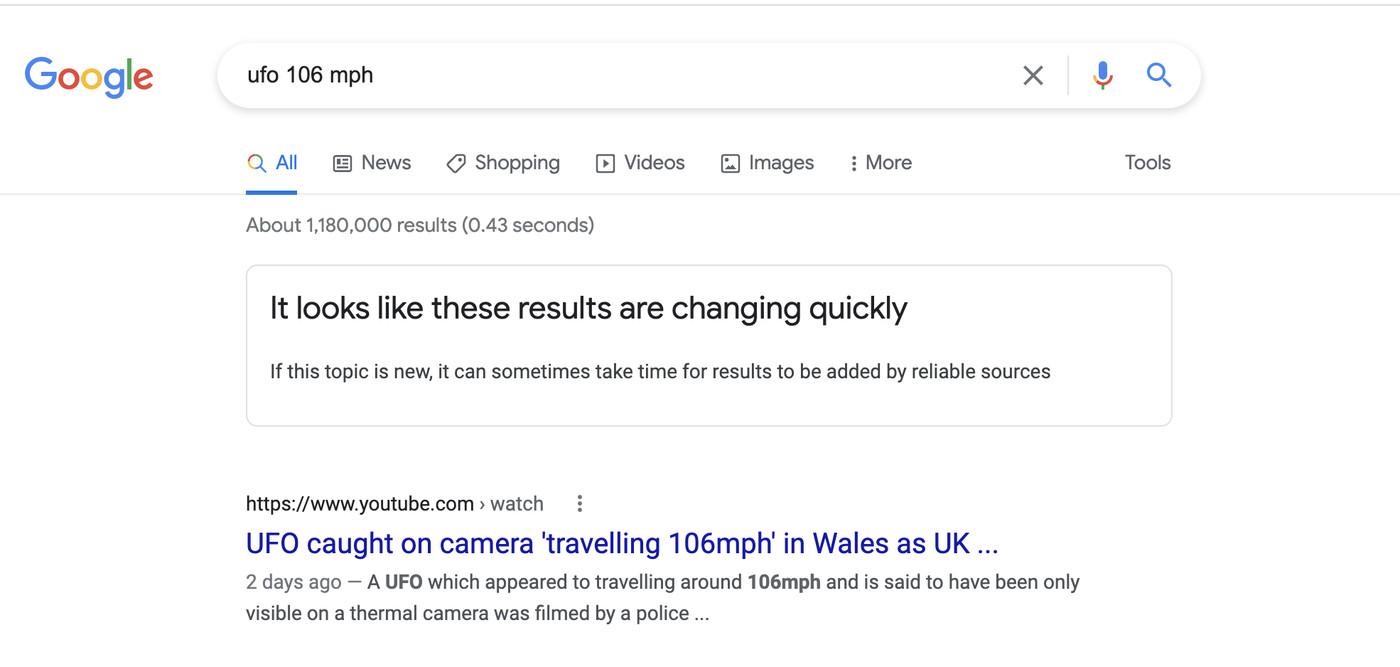}
  \caption{A Google warning banner for unreliable information in search results (Public domain, via Google Blogs~\cite{Sullivan2021}).}
  \Description{A Google warning banner for unreliable information in search results.}
  \label{fig:banner}
\end{figure}

Concerns about misinformation in search results led Google to place warning banners at the top of suspect search result pages. Figure~\ref{fig:banner} shows an example~\cite{Sullivan2021}. 
The warnings are intended to help people ``confidently evaluate the information [they] find online''~\cite{Sullivan2021}, and the initiative has been praised as a promising step toward protecting the public from unreliable information~\cite{Ghaffary2021}.

These search page banner warnings are distinct from traditional forms of fact-checking in which warning labels have been attached to specific pieces of content in social media, on feeds, and even search engines~\cite{Aslett2022}. Item-specific warnings have extensive prior literature, albeit with conflicting findings. Previous studies report null effects of warnings attached to individual news items and results on search pages~\cite{Aslett2022}, as well as backfire effects of warnings attached to individual social media posts~\cite{Pennycook2020}. For instance, Pennycook and colleagues~\cite{Pennycook2020} found that people who saw a list of headlines with warning labels attached to some automatically assumed that the headlines without these labels attached were accurate, the “Implied Truth Effect.”

More relatedly, research also addressed the effect of general information reliability warnings (such as the one shown above by Google). 
Evidence on general warnings’ efficacy is also mixed. Roozenbeek et al.~\cite{Roozenbeek2022} found that YouTube videos about hallmarks of misinformation improved viewers’ ability to discern accurate from inaccurate content, but other studies have found null effects of the general warning approach. For instance, Greene and Murphy~\cite{Greene2021} tested how warning banners alerting people that not all news stories are accurate impacted their recognition of false stories about Covid-19. They found that people who saw these warnings rated the false stories’ truthfulness no differently than those who saw no warning, regardless of whether the warnings used positive or negative framing~\cite{Greene2021}.

More concerningly, recent studies have found that general notices about misinformation can harm credibility perceptions of \textit{accurate} news~\cite{VanDerMeer2023, Clayton2019}. Van der Meer et al.~\cite{VanDerMeer2023} showed participants one of four general warnings (or no warning) before showing them four factual headlines about climate change. The authors found that a warning about the omnipresence of misinformation actually decreased participants’ credibility perceptions of the accurate headlines relative to all other conditions. Clayton et al.~\cite{Clayton2019} directly compared the effects of showing a similar general warning with specific warnings on people’s accuracy assessments of headlines about Donald Trump. These researchers found that specific corrections of false headlines had no adverse effects, but the general misinformation warning again decreased participants’ accuracy ratings of the true headlines.

These studies show the importance of understanding how warning banners affect users’ evaluations of both true and false content in search, and whether they also moderate the possible effects of rank and misinformation on click behavior and trust. Based on these studies of general warnings from other information-seeking contexts, we propose that:

\textit{\textbf{H5}: Ratings of results’ trustworthiness by people who see a warning banner will be lower than the ratings by people who see no warning.}

\section{Methods}
We used three online experiments to address the hypotheses listed above. The experiments were somewhat different in their design, though all had the same basic structure where participants interacted with a search engine result page, and were then asked to evaluate the trustworthiness of some of the results.

This research was approved and exempted from full review by our institution’s IRB office (protocol \#0010181). 
All hypotheses and analyses were pre-registered on AsPredicted.com. The preregistrations, code, anonymized datasets, and detailed results for each analysis are available at the \textbf{anonymous} OSF repository for this project (\url{https://osf.io/n2b3s/?view_only=0ae0c8d5d110493d933ab76a633fedd6}).

\subsection{Procedure}
Our experimental setup, across all three experiments, involved showing online participants three search engine result pages (SERPs) in a random order. 
We designed these pages to look like real pages of Google search results, with a query shown in a search bar and ten search results listed below it. In each experiment, two of these pages were decoys with queries about other topics, used for distraction and not used in the data analysis. 
The remaining query and result page was the setting for each experiment: it had a query about Covid-19 selected at random from three different queries for each experiment, and a set of associated results shown in random order.
The main factors of the Covid-19 search pages that we manipulated in every experiment were the order of the results with accurate information, and the presence and location of a result with misinformation, 
with a control group which saw ten accurate results presented in Google’s interface.
If the misinformation result was shown alongside the accurate results, it was positioned in one of the three top-ranked results on the page.

We prepared the different queries and the search results for each query in advance. 
We sourced the results shown on these experiment pages from Google. 
We collected the accurate results for each experiment by issuing the query on Google, then selecting results that came from established medical websites and agreed with the scientific consensus about the query at the time. 
The misinformation results shown were either gathered from Google or created by the researchers, as described below. 
All the queries, including the decoy queries, and their search results are included in the ``Stimuli'' document within the Supplementary Materials file. 

To avoid potential effects of a single query, a specific misinformation result, or a fixed ranking of the accurate results, we randomized all these factors. First, we randomly assigned participants in each experiment to see one of the experiment's three distinct queries about Covid-19. For the conditions in which misinformation was shown, we randomly selected one of the three misinformation results specific to that query, and showed this result in the designated rank. Lastly, we randomized the order of the accurate results in every condition across all experiments.

\begin{figure}[ht]
  \centering
  \includegraphics[width=\linewidth]{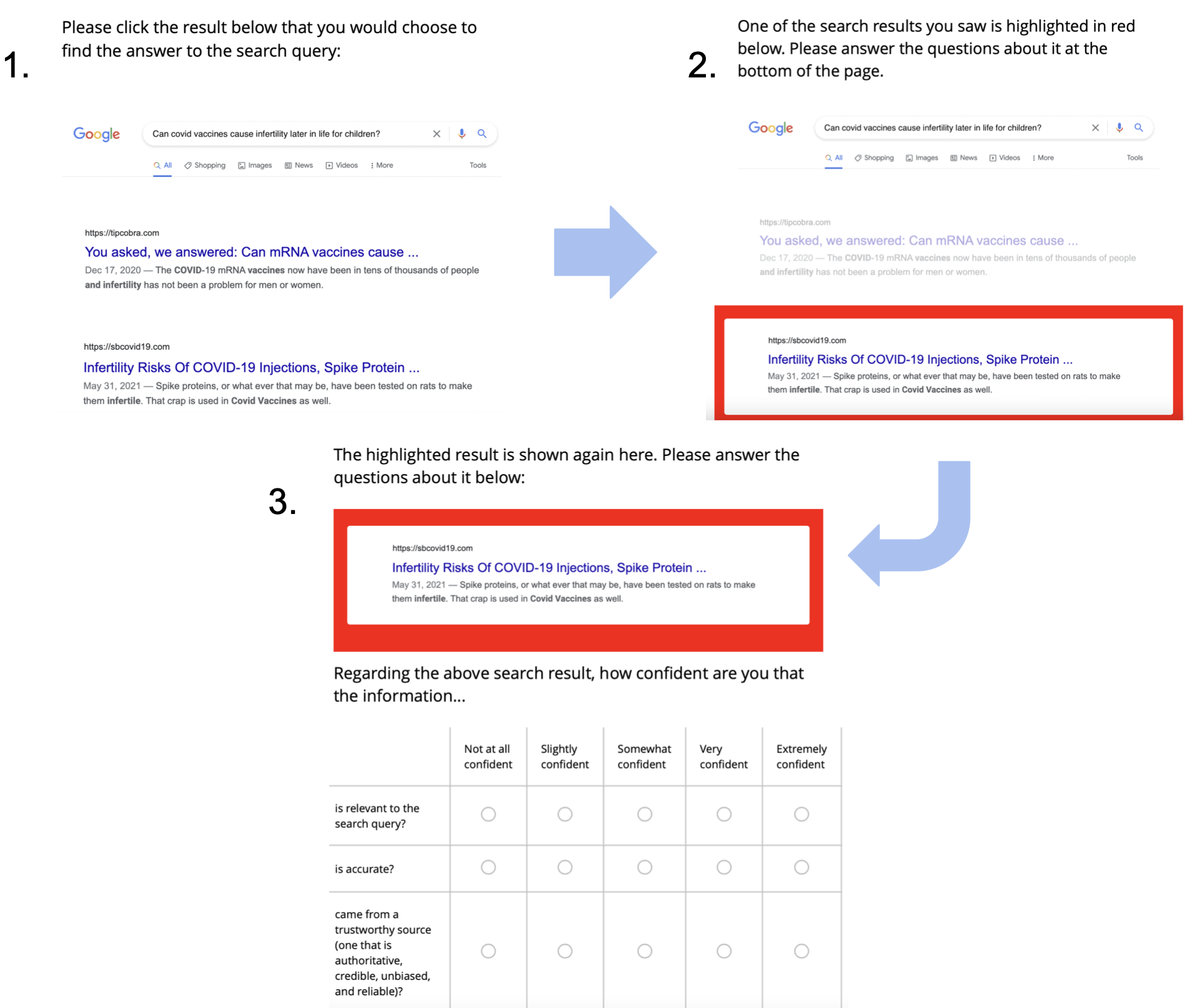}
  \caption{The survey interface developed for the three experiments. This figure shows an example of a page seen in Experiment~2 in the condition with misinformation in the second rank. (Note: there were ten total results shown on the first page that could not all be captured in a screenshot. We ensured that participants saw all ten results by making them scroll to the bottom of the page to continue the survey).}
  \label{fig:study}
  \Description{A series of screenshots from the survey in Experiment 2}
\end{figure}

For each experimental query, we measured the search action (the participant clicking on a result) and the participant's evaluation (rating of the trustworthiness of a result) with the following procedure.  
We asked participants to click on one result that they would choose to answer the query in question. 
After participants clicked a search result, we showed the page of results again, with one of the results highlighted in red and the rest of the results grayed out. 
We showed the highlighted result again at the bottom of the page, where we asked participants to indicate their opinions of the result’s trustworthiness (see detailed measures in Section~\ref{sec:measures}). 
The reason we showed the result in the original page context was to subtly remind the participants of its rank on the page. 
In each experiment, the result(s) participants rated were chosen randomly, after a balancing procedure to ensure we would have ratings for clicked and non-clicked results, as well as accurate and misinformation results. 
We also collected participants’ ratings of the result’s accuracy and relevance, 
which we only used to check their correlations with trustworthiness ratings. 
After rating the result(s), participants were directed to answer optional questions about their relevant attitudes and demographics (see Section~\ref{sec:measures}).
The survey procedure is illustrated in Figure~\ref{fig:study}.

We used Prolific Academic for recruitment in each study. We screened for participants who resided in the U.S. (due to the unique relevance of the Covid-19 vaccine queries to this population at the time of the study) and who were fluent in English (due to the search results all being in English). We also requested to get a sample evenly split between men and women.
Before starting the experiment, participants were asked to give their informed consent and explicitly confirm that they were using a desktop or laptop computer (which was necessary for the survey to work properly). 
At the end of the survey, we showed participants a debriefing statement before we confirmed their compensation. 
The statement disclosed that they may have seen information with harmful misinformation about Covid-19, and that they should not take the information from our search results as medical advice.

While all using the same basic structure, the three experiments had different designs, detailed next, which allowed us to test the various hypotheses.  

\subsection{Experiment 1}
\begin{figure}[ht]
  \centering
  \includegraphics[width=120mm]{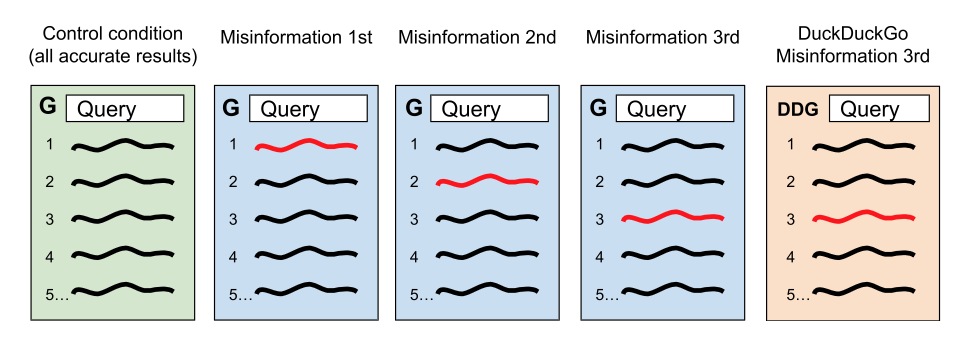}
  \caption{Conceptual illustration of the conditions in the between-subjects experimental design used in Experiment 1 and Experiment 2.}
  \label{fig:treatments}
  \Description{A diagram showing the conditions in the first two experiments.}
\end{figure}

In Experiment~1 we used five between-subject conditions, alternating the presence and rank of the misinformation result and the search engine the results were associated with.
The experiment design is illustrated in Figure~\ref{fig:treatments}. 
A control group showed all accurate results and displayed the search results page as if created by Google. The four treatment conditions showed misinformation ranked first, second, or third in Google, or ranked third in DuckDuckGo.

In Experiment~1, we used queries and associated results on the topic of the widely discussed safety of Covid-19 vaccines for children. 
The three queries we used for this experiment represented child-specific concerns about the vaccines’ safety, potential long-term risks to fertility, and potential cancer-causing effects.
These concerns have been documented frequently in search logs~\cite{Chang2023}. 
Although the topic remains politically controversial, the FDA had authorized the vaccines for use in children starting at age five and multiple peer-reviewed scientific reports supported their safety in this age group~\cite{Hause2021, Lv2021}. This data provided a reliable benchmark for identifying medical misinformation. 

We sourced misinformation results for our treatment groups in Experiment~1 by searching for each of the three queries about Covid-19 vaccine safety on Google and collecting results that 1) came from social media sites and 2) had extreme claims that the vaccines were unsafe. 
We validated our choices of misinformation results by running a pretest in which we showed pages of ten search results to a sample of individuals from varying socioeconomic and political backgrounds (N = 30). 
Five of these results had information that contradicted the scientific evidence showing Covid-19 vaccines were safe for children. 
We randomized the results’ order for each participant and asked them to select any results that they believed had false information. We chose the three results that were most frequently chosen as misinformation to be used in experiments~1 and~2. The ``Pretests'' document in the Supplementary Materials provides additional details about the pretest procedure.

Participants in Experiment~1 were asked to indicate the perceived trustworthiness of \textit{three} results on the search page. Participants first rated the result they had clicked, then rated two other non-clicked results. 
We randomly selected the non-clicked results from the top five results shown to the participant, due to past studies indicating that people pay the most attention to search results in this range~\cite{Hotchkiss2005}. 
Importantly, for participants in the misinformation treatment conditions, one of the results they were asked to evaluate was always the misinformation result, regardless of whether they clicked it. 
This design allowed us to measure participants’ levels of trust in a range of results: those clicked by participants, those which were not clicked, and results that had both accurate information or misinformation.

We recruited 1000 participants (200 per condition) for Experiment~1 to be able to detect a small expected effect size of f=0.1 for the effect of rank on trust in accurate results controlling for click status, at 80\% statistical power (the ``A Priori Power Analysis'' document in the Supplementary Materials provides additional details on this analysis). 
After collecting this sample, we excluded data from one participant who had a missing value for their unique ID and one who failed our attention check, leaving 998 participants and 2,994 trustworthiness evaluations for analysis. 
Our final sample for Experiment~1 consisted of young adults of average age of 37.2 years (SD=13.4); 47.9\% identified as women. 
Participants were left-leaning (66\% identified as leaning liberal, 26\% leaned conservative, and 8\% had no opinion), supportive of vaccination (85\% of participants indicated support for vaccination, 7\% indicated non-support, and 8\% indicated having no opinion). 
We discuss the challenges presentation by the skew of this sample in Section~\ref{sec:limitations}. 
Participants were generally trusting of the search engine they were shown (for Google, 78\% indicated trusting the search engine while 16\% distrusted it and 7\% indicated having no opinion; for DuckDuckGo, 79\% indicated trusting the search engine and the remaining participants indicated having no opinion).

\subsection{Experiment 2}
Experiment~2 expanded on Experiment~1 by testing the effect of rank and misinformation on trustworthiness when the sources of search results are not well-known to participants. 
To this end, we obtained eighteen candidate source URLs from real Google search results for the same query as Experiment~1. 
We then ran a pretest using a separate group of Prolific participants (N=66) to validate their unfamiliarity with these sources. We asked participants whether they recognized each website and how realistic each website sounded (to ensure that the sites we used would still seem plausible to participants in the main studies; the ``Pretests'' document in the Supplementary Materials provides additional details). 
We retained 13 sources that had at least 87.5\% of participants who failed to definitively recognize them. We used the same search results’ \textit{text snippets} from Experiment~1, but we randomly assigned one of the less-known sources to be associated with each of the results, instead of their original sources.
We saw evidence that participants noticed the uncertainty of the sources in that the average trustworthiness rating given to accurate results in this experiment ($\bar{x}$ = 2.81 scale points) was lower than in Experiment~1 ($\bar{x}$ = 3.41 scale points).\footnote{We do not use a statistical test to compare these scores since they were drawn from different experiments and samples.}

We changed the procedure from Experiment~1 to Experiment~2 in three ways: we revised the trustworthiness rating prompt, we asked participants to rate only \textit{one} of the search results, and we expanded the range of non-clicked results they could be assigned to rate to all results that appeared on the page. 
Instead of asking about trust in the results' information, we asked about trust in the results' sources to better represent validated measures of trustworthiness from past studies (see Section~\ref{sec:measures} for details).
We only asked participants to rate the trustworthiness of one result because we did not want to prime participants to pay extra attention to the manipulation of result rank, which could engender unnatural responses due to demand characteristics.
Using a similar approach to that used in Experiment~1, we obtained ratings of random results, varying in accuracy and click status, though since we asked for a single rating per participant, we only used between-subject variation. 

We again recruited 1000 individuals, and excluded two from analyses who failed our attention check, leaving a final sample of n = 998 participants. Our sample had similar characteristics to that of Experiment~1 (mean age = 40.2 years, SD age = 14.6 years, 48.5\% women; the ``Descriptive Statistics'' document in the Supplementary Materials provides additional details).

\subsection{Experiment 3}
We ran a third experiment to test how the outcomes from Experiment~1 and Experiment~2 might change when people are faced with uncertainty about not only the results' sources, but also about the answers to the queries. To that end, we replaced the vaccine safety queries from experiments~1 and~2 with three new Covid-19 queries that lacked medical consensus at the time of the study. We obtained uncertain queries by searching for questions in the news about Covid-19 and found nine that had at least two conflicting answers on the first result page, each endorsed by medical sources. We then conducted a pretest to confirm Prolific participants were uncertain about these queries. 
The pretest identified three queries, used in this experiment, for topics that participants were most split about: the possibility of animal-based transmission, the value of double-masking, and the efficacy of natural immunity compared to vaccine-induced immunity.
We saw evidence that participants noticed the uncertainty of the queries in that the average trustworthiness rating given to accurate results in this experiment ($\bar{x}$ = 2.55 scale points) was lower than in the former experiments. The ``Pretests'' document in the Supplementary Materials provides more details about this pretest. 

For each of these queries, we found ten authoritative search results from Google (defined as those having information supported by multiple medical sources). 
Due to the uncertainty of the new queries, we were unable to definitively classify any of the Google results as entirely accurate or as misinformation. 
For the misinformation manipulation, we created several fake results with blatant false claims about the pandemic in general (e.g. “Covid is a hoax”). To ensure that people perceived these results as being false but still realistic, we did another pretest on a separate group of Prolific participants (N=80) in which we asked them to rate the misinformation results we created and some of the accurate results on perceived accuracy and perceived likelihood of appearing online (the ``Pretests'' document in the Supplementary Materials provides additional details). 
We ultimately selected three misinformation results for each query that had significantly lower accuracy ratings and were statistically indistinguishable in perceived likelihood of appearing online. 
Finally, when displaying the result pages, we randomly paired each search result’s text snippet with one of the unknown sources like in Experiment~2.

\begin{figure}[ht]
  \centering
  \includegraphics[width=\linewidth]{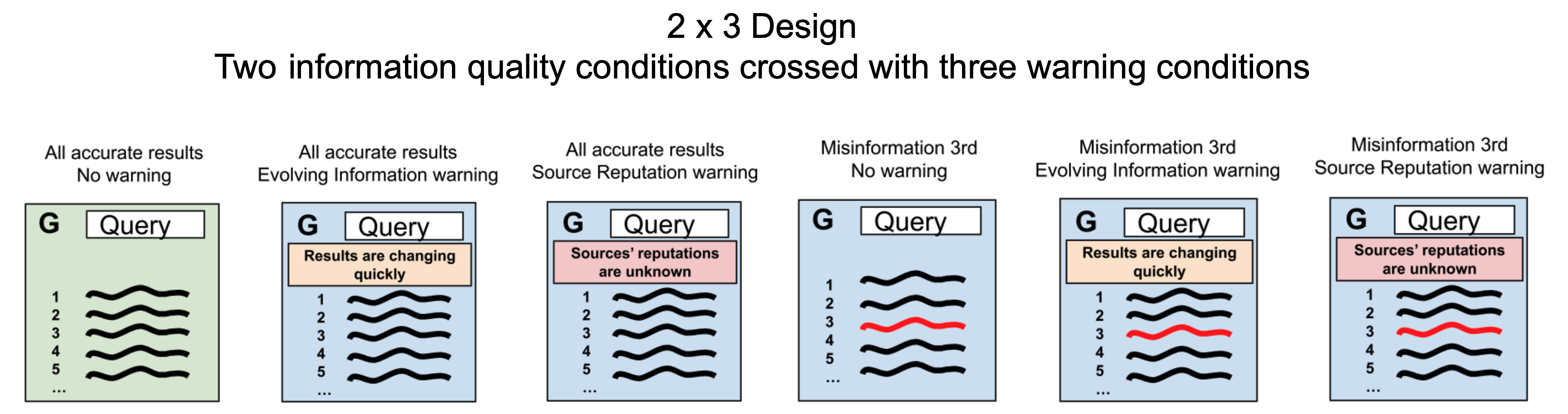}
  \caption{Conceptual illustration of the conditions in the between-subjects experimental design used in Experiment~3.}
  \Description{A diagram showing the conditions in the third experiment.}
  \label{fig:ex3treatment}
\end{figure}

We made three changes to the experimental design for this study. 
Instead of showing a misinformation result in one of the top three ranks, we only showed it in the third rank due to the fact that misinformation’s exact rank was not a moderator of any effects seen in the previous studies. For the same reason, we omitted the DuckDuckGo condition from this study. 
Most importantly, we added a new between-subjects manipulation in which we displayed one of two possible warning banners, or no banner, at the top of the search page, to test the effect of banners on the evaluation of results. 
This two-by-three design is illustrated in Figure~\ref{fig:ex3treatment}. 

\begin{figure}[ht]
  \centering
  \includegraphics[width=\linewidth]{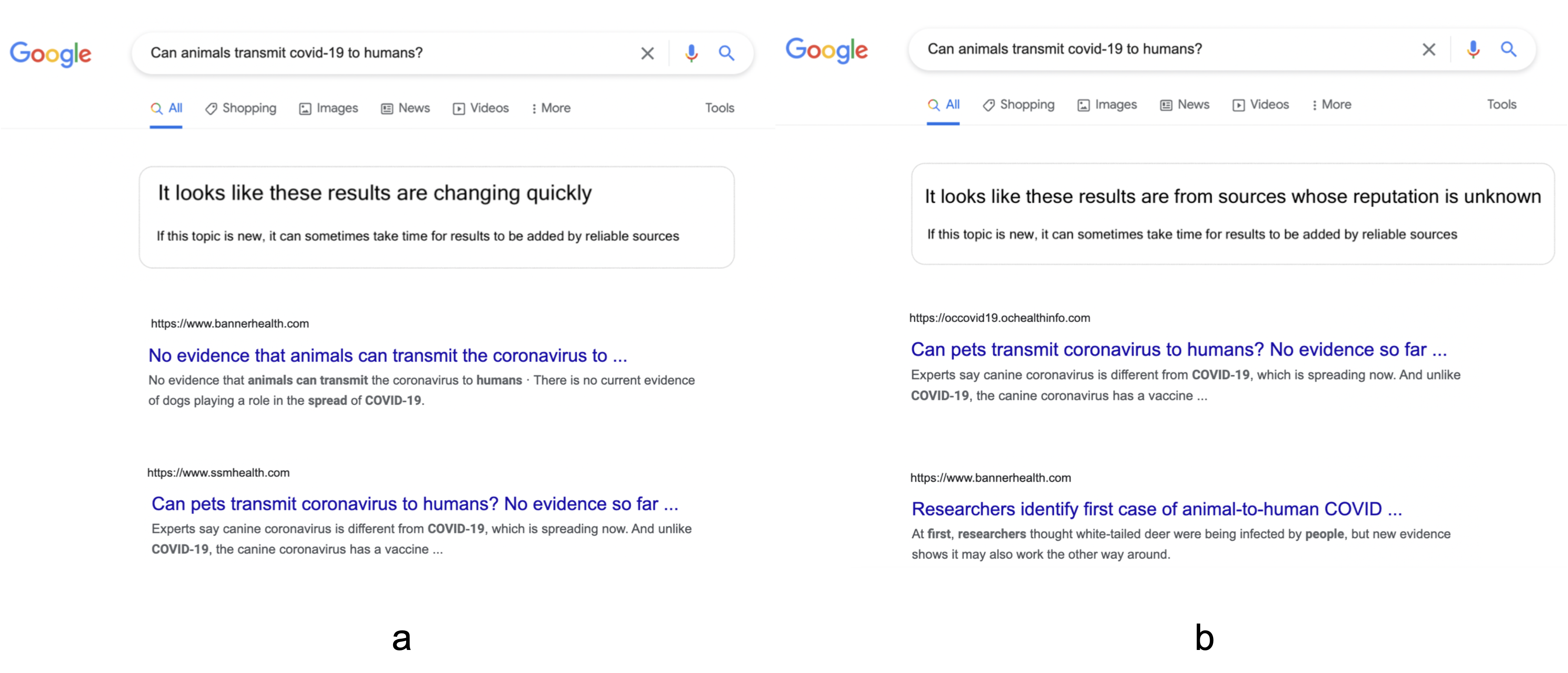}
  \caption{The two types of warning banners in Experiment 3. On the left,  Google's ``evolving information'' warning. On the right, the ``source reputation'' warning we created.}
  \Description{Examples of two warning banners that served as treatments in Experiment 3.}
  \label{ex3warnings}
\end{figure}

One of the banners we implemented, which employs a design that is currently used by Google, mentions that the results are changing quickly with subtext telling participants that it can take time for results from reliable sources to appear (the “evolving information” warning). 
The second banner has the same subtext as the ``evolving information'' warning, but its main text instead says that the sources' reputations are unknown. This ``source reputation'' warning tests whether alerting participants to the unreliability of search results’ sources has different effects on trustworthiness evaluations than warning them about the unreliability of the information itself.
The two banners are shown, as they were displayed to participants, in Figure~\ref{ex3warnings}.

We recruited 1200 participants (200 per condition as recommended by the power analysis). We included all participants' data in our analyses because none failed our attention check. Experiment~3's sample closely resembled those of experiments~1 and~2 with a mean age of 36.2 years (SD=12.9) and 48.8\% identifying as women. The ``Descriptive Statistics'' document in the Supplementary Materials provides additional details about the sample.

\subsection{Measures}
\label{sec:measures}
Our hypotheses test the potential effects of rank, misinformation presence, and misinformation warning banners on our two main dependent measures: click behavior and perceived trustworthiness.
Participants were first asked to click a result on the search page with these instructions: “Please click the result that you would choose to find out the answer to the question being searched.” Participants were only allowed to click one result, and we recorded the rank of this result as well as whether it contained accurate information or misinformation.

The measurement of perceived trustworthiness was more complex.
We had participants rate how trustworthy they perceived the result(s) to be by asking them “How confident are you that the result is trustworthy?” in Experiment~1, or “How confident are you that the information came from a trustworthy source (one that is authoritative, credible, unbiased, and reliable)?” in experiments~2 and~3. 
We modified the prompt after Experiment~1 to better disambiguate trustworthiness from accuracy and other constructs. 
The terms used in experiments~2 and~3 represent factors documented by Metzger and Flanagin~\cite{Metzger2013} that contribute to the perceived credibility of online information. 
We found that adding these terms decreased the correlation between accuracy and trustworthiness ratings (from r = 0.93 in Experiment~1 to r = 0.81 in Experiment~2 and r = 0.84 in Experiment~3), suggesting that this revision helped differentiate these factors.

Participants responded using a Likert scale of confidence (1=“Not at all confident”, 5=“Extremely confident”). This scale has been used to measure the strength of people’s attitudes in multiple contexts~\cite{Cadario2021, Hall2017}, including attitudes about the trustworthiness of Covid-19 information~\cite{Thorpe2022}. Although formal models of perceived trustworthiness break the construct down into separate antecedents, we used one item to prevent participants from experiencing fatigue. 
This decision was supported by an initial pilot study in which we used five separate items to measure participants’ trust toward the search engine results, where these items had high internal consistency (N=102; Cronbach’s alpha=0.93), as well as other studies that have found similar internal consistency for multi-item trustworthiness measures~\cite{Haas2017, Ma2017} and those that have only used one item to measure people’s trust in information~\cite{Pennycook2019}.

Finally, to better understand the demographic and attitudinal composition of our samples, we measured the following variables in a post-task survey: trust in the search engine shown in the study~\cite{Ma2017, Mayer1995}, support for vaccination~\cite{Akel2021}, political ideology~\cite{Jakesch2019}, age, and gender. We report descriptive statistics based on these factors in the “Participants” subsections above for each experiment. The ``Descriptive Statistics'' document in the Supplementary Materials provides breakdowns of these variables for each sample.

\subsection{Analytic Approach}
We used logistic regression models to analyze the effects of rank, misinformation, and warning banners on click behavior, and linear regression models to analyze these factors' effects on trustworthiness ratings. 
We verified that our data satisfied the statistical assumptions of these models before using them. 
We used the Type III Sums of Squares ANOVA to test for interactions; if no interactions were significant, we reran the model using the Type II setting to test for main effects. 
To decompose significant effects, we used pairwise comparisons of estimated marginal means for categorical predictors, and simple slopes analysis for continuous predictors, applying Bonferroni adjustments to each model's p-values.
All analyses were done with RStudio version 2022.07.0 and are reproducible using the data and code in the OSF repository for this project. 

The analyses presented below deviate from the preregistered approaches in three minor ways. First, we preregistered using a linear regression to model the effect of rank on results' click probability, but later corrected the approach to using, as appropriate for a binary outcome, a logistic regression model. 
Second, we preregistered testing whether misinformation’s presence impacts clicks or trust toward accurate results \textit{above and} below it; however, the literature suggests that misinformation’s presence will most likely affect impressions of an accurate result immediately below it, so we focused the hypothesis and report on models considering the below results specifically. 
Third, we preregistered models testing for 3-way interactions between results' rank, click status, and information accuracy when predicting trustworthiness ratings. Because there was an unexpectedly low number of participants who clicked on the misinformation result, we lacked statistical power to detect this type of interaction if it exists; thus, we report on models in which we test whether rank interacts with click status or accuracy individually. 
Importantly, none of the findings reported below changed when running the preregistered analyses, except for one secondary analysis on differences in the rank-click relationship depending on warning condition (see “Responses to Warnings” below). 
The folder called “Detailed Analysis Results” on the project’s OSF repository provides a detailed comparison of the preregistered and reported models' outcomes.

\section{Results}

We present the analyses for click behavior and trustworthiness judgments in response to the three main factors we manipulated: result rank, the presence of misinformation, and the inclusion of a warning banner about unreliable information or sources. 
Due to the similarity between the three experiments and the increased statistical power gained from aggregating their data, we mostly report on combined estimates for all the data below, while showing each experiment's outcome in the corresponding figures. The ``Detailed Analysis Results'' folder on the project’s OSF repository provides experiment-specific results.

\subsection{Responses to Rank}
\begin{figure}[ht]
  \centering
  \includegraphics[width=120mm]{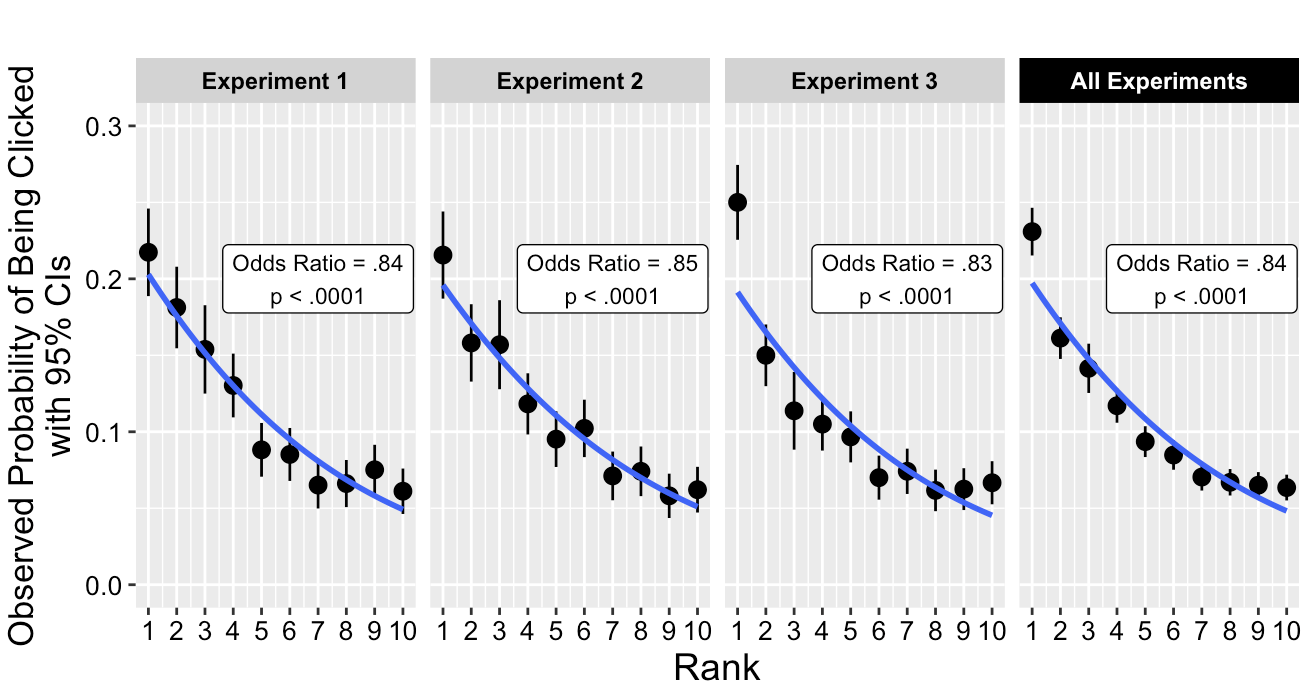}
  \caption{
  \textbf{Higher rank~(X-axis) is associated with an increase in an accurate result's click likelihood~(Y-axis).} 
  This relationship (blue line) is stable across experiments~(columns). 
  The blue lines are fitted by the logistic regression model and show the predicted probabilities of each rank being clicked.
  Error bars represent 95\% Confidence Intervals of the sample proportions. 
  }
  
  \Description{A visualization of the decreasing tendency to click on accurate search results as they get ranked lower on the page.}
  \label{fig:rank_click}
\end{figure}

Across all three experiments, our data supported~H0: accurate results in higher ranks were more likely to be clicked. Figure~\ref{fig:rank_click} provides a detailed overview of the rank-click relationships for each of the experiments (three leftmost panels), and in aggregate (rightmost panel). 
The X-axis represents the rank of a result, and the Y-axis is the likelihood of an accurate result at this rank to be clicked. 
For example, the left-most point in the left-most panel (``Experiment 1'') shows that an accurate result appearing at the highest rank in this experiment had a 22\% likelihood to be clicked. 
Each datapoint
represents probabilities generated from 199 to 402 individuals' click decisions, depending on the experiment(s) and rank displayed (since not all participants were exposed to accurate information in the first three ranks).
The figure demonstrates that there is a significant relationship between rank and click likelihood for each experiment: for every drop in position by 1, the odds of an accurate result being clicked decrease by 16\% on average in Experiment~1, 15\% in Experiment~2, 17\% in Experiment~3, and 16\% in the aggregate data~(all p's < .0001). This effect was robust when our models controlled for the presence and rank of misinformation on the page and the specific search engine the results were attributed to in Experiments~1 and~2 (though the strength of this relationship differed based on which type of warning was shown: see Section~\ref{sec:warnings}). 

We also found support for~H1, in that participants trusted accurate results they clicked more than accurate results they did not click by 
0.69 scale points~(F(1, 4008) = 357.45, p < .001) in the aggregate data. This effect was stable when our models controlled for the rated result's rank, its accuracy, the presence and rank of misinformation on the page, and the specific search engine the results were attributed to in Experiments~1 and~2.

\begin{figure}[ht]
  \centering
  \includegraphics[width=\linewidth]{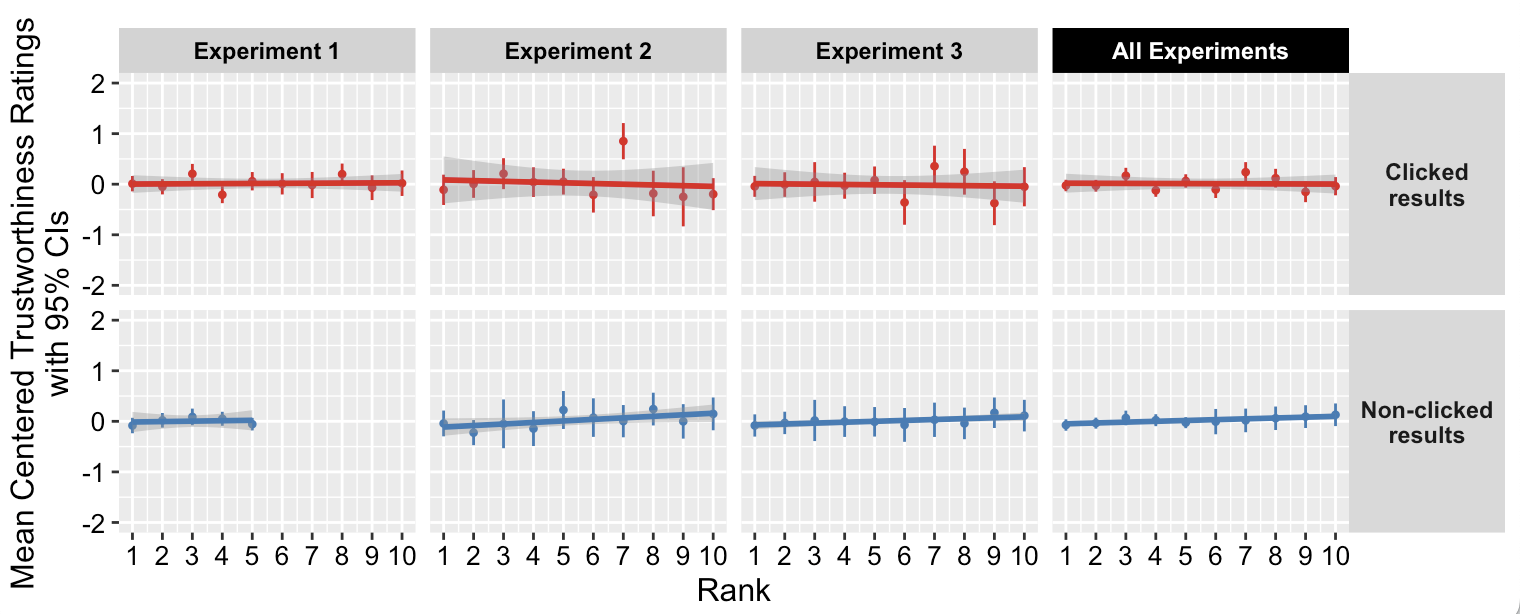}
  \caption{\textbf{Rank~(X-axis) does not affect the evaluation of trustworthiness~(Y-axis) of accurate results} even with increased statistical power from aggregating the data across the three studies.
  This lack of relationship is robust across experiments~(columns) and for clicked results (top row, red) as well as non-clicked results (bottom row, blue). 
  The trend lines represent the predicted change in trustworthiness ratings per unit decrease in rank fitted by the linear regression models.
  Error ranges represent 95\% confidence intervals of the centered trust ratings within the rank and click category for each experiment.
  }
  \Description{A visualization of the (flat) relationship between an accurate result's rank and its perceived trustworthiness.}
  \label{fig:rank_trust}
\end{figure}

Given the effect of rank on click probability, and the relationship between clicking and trustworthiness evaluations, it appeared plausible that rank impacts the perceived trustworthiness of results on search pages~(H2). 
However, that is not what we found. Unlike click behavior, evaluations of trustworthiness did \textit{not} vary with the rank of accurate search results in linear regression models, leaving H2 unsupported. 
Figure~\ref{fig:rank_trust} shows that the lack of association between rank and trust was robust for both clicked and non-clicked results across all three experiments (representing varying levels of uncertainty associated with sources and queries) and when aggregating their data for increased statistical power (rightmost panel).
The X-axis is the rank of each result, and the Y-axis is the average of the mean-centered trustworthiness ratings participants gave to results at that rank.\footnote{See the ``Detailed Analysis Results'' folder on the OSF repository for an explanation of why we mean-centered the results for the purpose of this figure, and only this figure.}  
The figure separates ranking for clicked (top) and non-clicked (bottom) results to show that the (non-)effect is robust for both types of results.
Each dot represents 19 to 431
trustworthiness evaluations, depending on the rank, experiment(s), and click category shown. For instance, the third point from the left in Experiment~1's top panel (which represents 92 ratings) shows that participants who clicked an accurate result in the third rank on the page gave it trustworthiness scores that were about 0.25 scale points higher than the average trustworthiness rating that participants gave the results they clicked across all ranks in this experiment. 
Altogether, Figure~\ref{fig:rank_trust} demonstrates that there is no significant association between the rank of an accurate result in search and how trustworthy people perceive its information or source to be.
This (non-)effect was robust when our models controlled for the presence and rank of misinformation, the presence and type of warning banner in Experiment~3, and the specific search engine the results were attributed to in Experiments~1 and~2. 

\subsection{Responses to Misinformation}
Responses to misinformation in the results page were nearly identical across the three experiments: while participants tended to not click and not trust misinformation, the presence of misinformation did not impact clicking or trusting accurate information below it.

First, misinformation was rarely clicked and highly distrusted. Across all ranks and experiments, only 2.6\% of participants who were exposed to inaccurate results clicked on these results. Misinformation results were clicked at a significantly lower click rate than accurate information displayed in the same position (rank 1: $\chi^2$(1, N=1996)=68.47, p < 0.001; rank 2: $\chi^2$(1, N=1996)=46.16, p < 0.001; rank 3: $\chi^2$(1, N=3196)=153.96, p < 0.001). Thus, H3a was supported. 
As expected in light of the trust-click association, participants also did not trust search results containing inaccurate information: participants trusted misinformation results less than accurate results by 1.56 scale points across all experiments (F(1, 5190) = 1696.6, p < .0001), supporting H3b.

Despite the clear responses to misinformation itself, the presence of misinformation in a search result did not affect participants’ propensity to click 
accurate information located immediately below it (as analyzed with a chi-square test). H4a is thus unsupported.

\begin{figure}[ht]
  \centering
  \includegraphics[width=\linewidth]{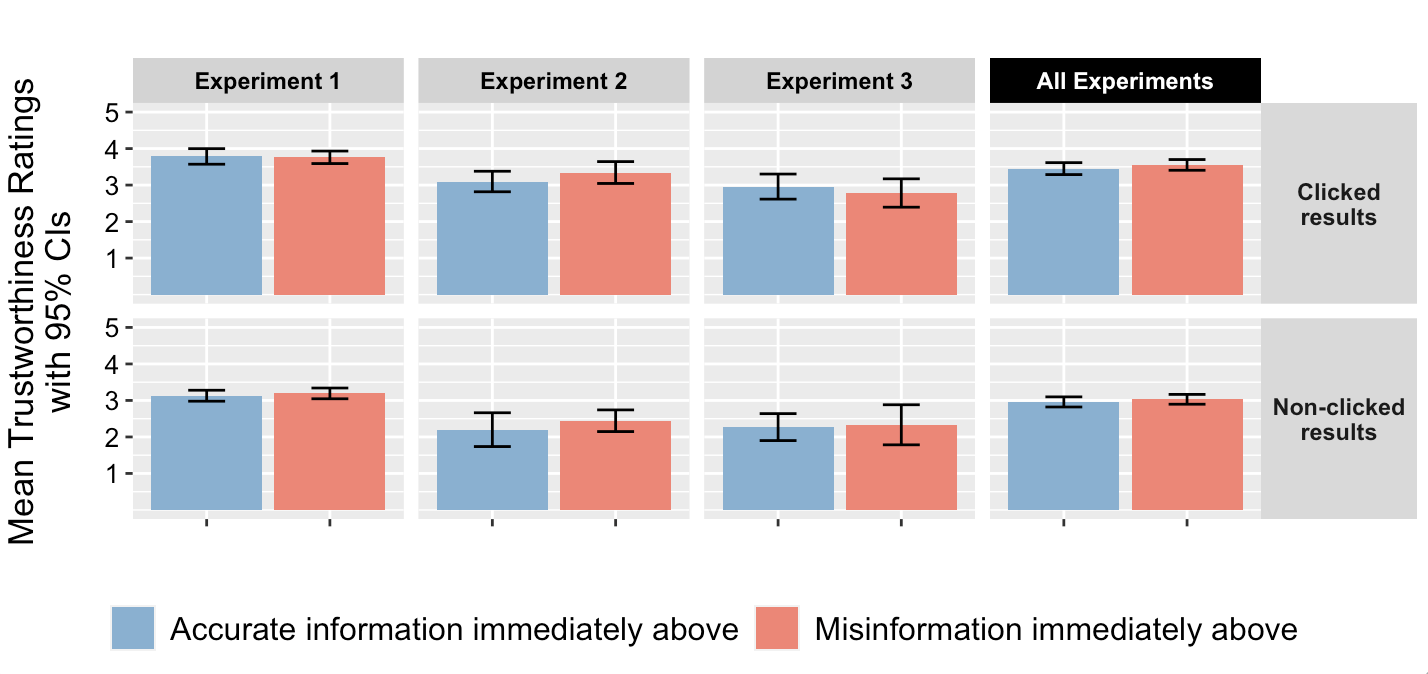}
  \caption{\textbf{The presence of misinformation~(red bars) had no effect on trust~(Y-axis) in an accurate search result appearing below it compared to when none of the results on the page contained misinformation~(blue bars), even with increased statistical power from aggregating the data across the three studies.} This null effect was stable across experiments~(columns) and for both clicked~(top row) and non-clicked~(bottom row) results. Error ranges represent 95\% Confidence Intervals of the means. 
  }
  \Description{A visualization of the nonexistent relationship between the presence of a misinformation result and the perceived trustworthiness of an accurate result immediately below it.}
  \label{fig:misinfo_impact}
\end{figure}

The presence of misinformation also did not shift the trustworthiness evaluation of the accurate information immediately below it in the linear models, as shown in Figure~\ref{fig:misinfo_impact}. 
The Y-axis displays the mean trustworthiness ratings given to accurate results in ranks 2, 3, and 4 by participants who saw a misinformation result immediately above these results (the red bars on the right side of each panel) compared to participants who saw no misinformation result anywhere on the page (the blue bars on the left).
Each bar in the figure represents   
trustworthiness ratings from 18 to 297 
participants, depending on the experiment(s) being displayed. 
We show results separately for evaluations of clicked- and not-clicked results. 
For example, the red bar in the bottom panel in the ``Experiment~2'' column shows that when participants were evaluating an accurate result they had \textit{not} clicked and which had been ranked immediately below a misinformation result, they gave it an average trustworthiness rating of two and a half, and the average for these results when there was no misinformation present (blue bar) was similar. 
This figure demonstrates that H4b is also unsupported: there is no significant difference in how trustworthy people perceive accurate results to be when a search page shows misinformation immediately above.
This (non-)effect was robust when our models controlled for the presence and rank of misinformation, the presence and type of warning banner in Experiment~3, and the specific search engine the results were attributed to in Experiments~1 and~2. 

\subsection{Responses to Warnings}
\label{sec:warnings}
Experiment 3 introduced an information reliability warning banner at the top of the result page. We performed a between-subject test of two different warnings (shown in Figure~\ref{ex3warnings}): the ``evolving results'' warning used by Google that says ``the results are changing quickly'' and a version we created called the ``source reputation'' warning. 

\begin{figure}[ht]
  \centering
  \includegraphics[width=120mm]{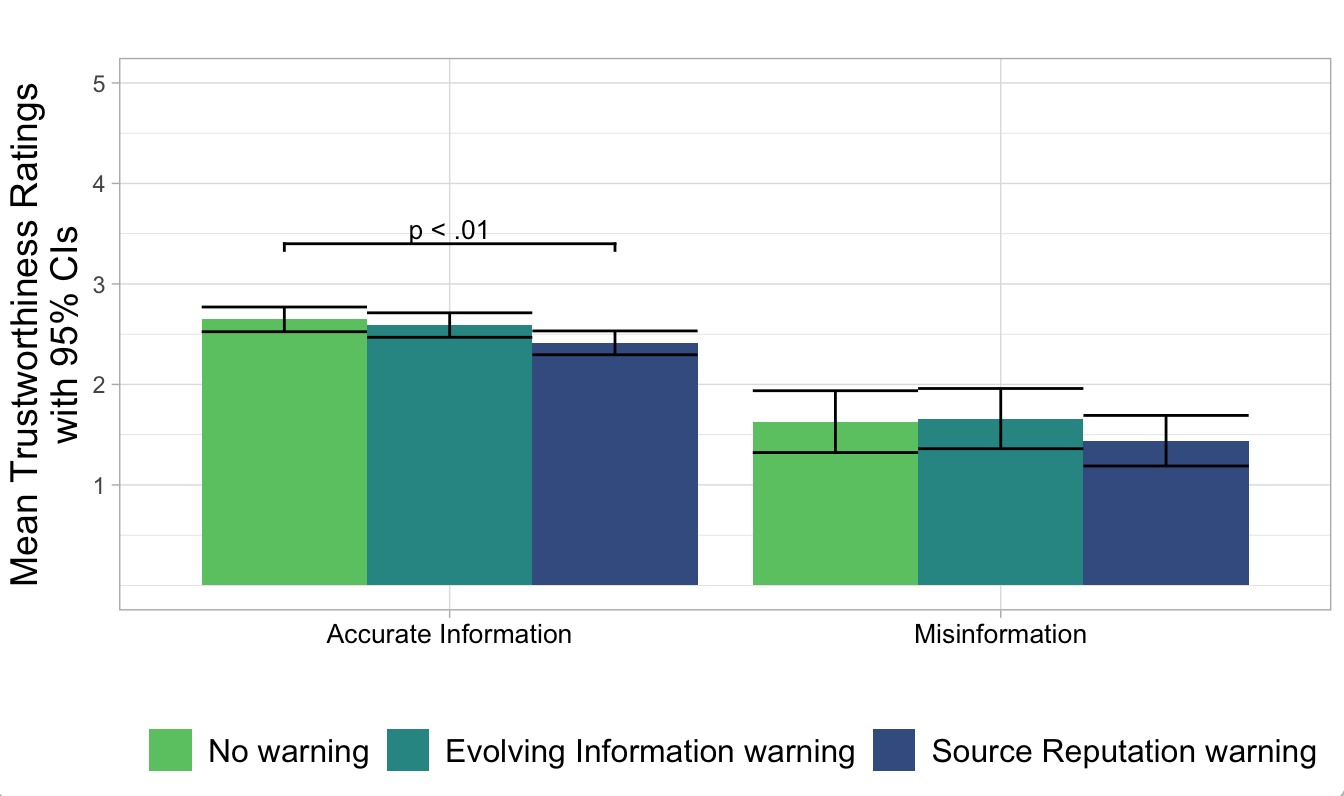}
  \caption{\textbf{The ``source reputation'' warning decreases trust in accurate information by about a quarter of a scale point}. 
  The bars show mean trustworthiness ratings (Y-axis) for accurate and misinformation results depending on the warning condition~(colored bars). Error ranges represent 95\% Confidence Intervals of the means.} 
  \Description{A visualization showing that the ``source reputation'' warning decreases trustworthiness perceptions of accurate results, but not misinformation.}
  \label{fig:warning_trust}
\end{figure}

We find that the ``source reputation'' warning affects both overall trust in accurate results and the strength of the previously reported rank-click relationship for these results, but Google's ``evolving results'' warning had no impact. 
Figure~\ref{fig:warning_trust} reports the effect of warnings on mean trustworthiness ratings of results, for accurate results (on the left) and misinformation results (on the right). 
The Y-axis shows the average trustworthiness ratings given by participants to accurate results~(left) and misinformation results~(right) when they saw one of the following warning conditions: no warning~(light green bars on the left), the ``evolving information'' warning~(dark green bars in the middle), or the ``source reputation'' warning~(dark blue bars on the right). 
Each bar 
in the figure represents 50 to 352 trustworthiness ratings depending on the type of result and the warning condition. 
The accurate information bars (on the left) 
show that the control group participants (leftmost bar) rated accurate results higher by 0.23 points on a five-point trust scale (F(2, 1195) = 4.68, p < .05) compared to those who saw the ``source reputation'' warning (third bar from left). This effect was robust when our model controlled for whether misinformation was present in the third result, whether participants were rating results they had clicked or not, and the rank of the result being rated (which had no effect on trust, as seen previously).
In contrast, the ``evolving information'' warning (second) had no effect on trust levels in accurate information compared to the control.
The right side of the figure shows that neither warning had a significant effect on (already low) trust levels in misinformation. 
The findings reveal a potential backfire effect: participants responded to the ``source reputation'' warning by showing lower trust in the accurate results, but the warning had no effect on trust in results containing misinformation. This 
lack of impact could be due to a floor effect, since trust in misinformation was already low at the baseline. See Section~\ref{sec:limitations} for further discussion.  
Given that one of the warning banners we tested decreased trust in the accurate information, our data provide partial support for H5.

This finding raises two concerns. First, the decreased trust in accurate medical information caused by the ``source reputation'' warning could theoretically steer people toward trusting and clicking misinformation instead. Second, it could cause people to rely on other cues in the search environment (such as rank) more heavily for their trustworthiness evaluations and click decisions. 

Our data do not support these concerns, though we do show that these types of warnings may still affect people's click behavior. Across warning conditions, misinformation’s trust ratings and click rates remain equally low, and misinformation still has no spillover effects on clicks or trust in accurate information below it.
Moreover, neither warning changed the (null) effect of rank on trust. 
Finally, while an exploratory analysis shows that banner warnings may have an impact on how people click on results based on their rank, the effect is opposite the expected direction: the ``source reputation'' warning decreases people's reliance on high rank when deciding which result to click. 
A logistic regression model of click likelihood showed a significant interaction between rank and warning condition. 
Participants who saw the ``source reputation'' warning were actually less likely to click on high-ranked accurate results compared to participants who saw the other warning (as observed through post-hoc pairwise contrasts of each warning condition’s rank-click relationship); however, neither warning condition showed significant differences in this relationship relative to the no-warning control. 

\begin{figure}[ht]
  \centering
  \includegraphics[width=90mm]{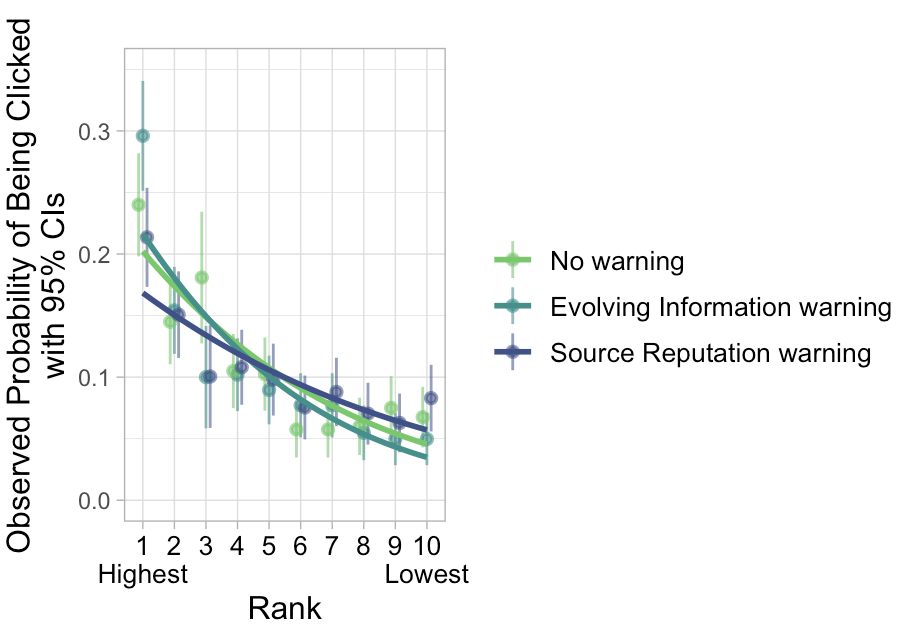}
  \caption{\textbf{The effect of warnings on the propensity to click higher-ranked results}. The ``source reputation'' warning~(dark blue line) weakened the effect of rank~(X-axis) on participants’ likelihood to click accurate results~(Y-axis) compared to identical search results with the ``evolving information'' warning~(dark green line). 
  Error bars represent 95\% Confidence Intervals of the proportions of clicks on each rank.}
  \Description{Warnings change the strength of the rank-click relationship}
  \label{fig:warning_rank_click}
\end{figure}

Figure~\ref{fig:warning_rank_click} illustrates the potentially changed relationship of rank and clicking in response to the warnings. 
The figure shows the rank of results (X-axis) and the likelihood of participants clicking on them (Y-axis) for the three warning conditions, with lines fitted by the logistic regression model showing the predicted probabilities of each rank being clicked within each warning condition. 
For instance, the dark green point at rank 1 shows that the highest-ranked result had about a 30\% chance of being clicked by participants in the ``evolving information'' warning group. 
Each datapoint represents probabilities generated from 199 to 402 individuals' click decisions, depending on the warning and misinformation presence conditions. 
The most moderate slope in the figure is that of the ``source reputation'' warning (darkest/blue line), which was associated with a 14\% decrease in odds of clicking per one-step drop in rank (N=398, odds ratio = 0.86, 95\% CI [0.82, 0.91], p < .0001), whereas the ``evolving information'' warning was associated with a 21\% decrease (N-402, odds ratio = 0.79, 95\% CI [0.74, 0.84], p < .0001). 
While the differences between these two slopes was significant, neither of these estimates significantly differed from the ``no warning'' group's 17\% decrease (N=400, odds ratio = 0.84, p < .0001). 
The ``Detailed Analysis Results'' folder on the project’s OSF repository
provides a more detailed comparison of these conditions, including our preregistered analysis which we slightly deviate from here.
These results show that warnings may accentuate, or moderate, the effect of rank of clicking behavior, causing people to rely more (or less) on top-ranked results. 

\section{Discussion}
\label{sec:discussion}
The foundational paper called ``In Google We Trust: Users' Decisions on Rank, Position, and Relevance''~\cite{Pan2007} has been widely misinterpreted, such as when it was cited by Epstein and Robertson~\cite{Epstein2015} as showing that ``users trust and choose higher-ranked results more than lower-ranked results''~(p. 4152). 
Instead, the original study actually showed that users trust Google that highly-ranked results would be the most \textit{relevant} to the search query. 
In other words, earlier work showed that individuals trust Google’s \textit{ranking}, but that this did not directly reflect on trust in the individual \textit{results}.

In this work, 
by directly measuring the perceived trustworthiness of individual search results, we showed that despite earlier conjectures~\cite{Metaxa2019, Epstein2015},
peoples’ trustworthiness evaluations are not affected by the rank of individual results. 
These null effects remained even when results on the search page were associated with unfamiliar sources; when the results page included misinformation results; when the query topics (and results) were lacking medical consensus; and in the presence of warning banners about unreliable information or sources on the result page.

The fact that rank and trust are not related has practical and theoretical implications. On the practical side, people rely on search results for critical information, including crisis events and medical advice~\cite{Chang2023, Song2022, Pogacar2017, palen2011supporting}. 
The absence of an association between rank and trust suggests that the reliance on rank for click decisions may be due largely to widespread awareness that rank corresponds to relevance, not trustworthiness. 
If people were to infer the trustworthiness of information from its position on the results page, they could become reluctant to take information from authorities (like the CDC in the case of Covid-19) if these authorities' results were not in the topmost ranks. 

At the same time, people's lack of response to rank when evaluating trustworthiness in search poses challenges to search engines' development of trustworthiness signals. Google and other companies put a lot of effort into helping users evaluate the individual results on a search page~\cite{Sullivan2021}. 
Showing---as we have here---that the rank ``signal'' may not impact users' evaluation of the content has important implications for the design of search engine interventions and ranking algorithms. 
For example, such designers may consider emphasizing (or developing) other measures that can help with the evaluation of trustworthiness for individual results, to help users make informed decisions. 

These findings also have important implications for research on search ranking. 
Our results raise the question of whether trustworthiness signals on search result pages can ever override the ``cognitive miser'' approach to search~\cite{Azzopardi2021, Operario2002}, 
in which users select high-ranked information because they blindly trust it to be most relevant (``In Google we Trust'')~\cite{Pan2007}. 
The research community needs to better understand the \textit{algorithmic} relationship of actual---not perceived---trustworthiness to ranking
and warnings. 
Some related work has been done in the context of search engine audits, focusing on the presence of misinformation in search pages~\cite{Zade2022, Juneja2021, White2014}.
Assuming the relationship between the trustworthiness of web sources and search engine ranking is not entirely robust, we need to understand what interventions and signals will help people better \textit{evaluate} trustworthiness, and reduce reliance on rank.

Indeed, our work provided some tentative evidence that the relationship between rank and clicking can be manipulated with trust-related design interventions. 
Although our ``source reputation'' warning decerased trust in the accurate results on average (more on that below), our analysis of the rank-click relationship under different warning conditions (Section~\ref{sec:warnings}) suggests a more nuanced effect on people’s search behavior. 
People exposed to the ``source reputation'' warning showed an increased likelihood of clicking for lower-ranked results compared to the ``evolving information'' warning, suggesting that the warning caused people to look further down the search result page, perhaps reducing reliance on top-ranked results. 
Thus, the ``source reputation'' warning may have served as a double-edged sword: the decreased trust in the presented information may encourage people to look at more results on the page, which could help people gain a more informed understanding of the topic. While past studies have similarly found that general misinformation warnings can decrease people's truthfulness perceptions of accurate information~\cite{VanDerMeer2023,Greene2021,Clayton2019}, other studies have found that these warnings can not only improve trustworthiness discernment, but can also provide other benefits, such as improving people's recognition of common manipulation techniques on social media~\cite{Roozenbeek2022}.
In sum, our findings show that the effects of these nonspecific warnings are multifaceted and warrant more research, particularly in search engines.

Our findings also add to the literature on the potential effects of misinformation and the approaches to address it. 
First, some potentially good news: 
while evidence from other domains suggested that exposure to misinformation could undermine people’s trust in accurate information~\cite{VanDerMeer2023, Song2022, Rapp2018}, misinformation in our study had no negative effects on the propensity to click or trust accurate results that followed. 
Participants did \textit{notice} the misinformation: they were less likely to click or trust results that contained misinformation in the experiment. 
Nevertheless, the presence of misinformation did not shift their evaluation of content ranked lower on the page.\footnote{Note that, in our experiments, we only reported on the difference of evaluation of the result immediately below the misinformation result; but the OSF folder called ``Detailed Analysis Results'' includes models that looked at the impact of misinformation presence on results in all positions on the page (e.g. higher-ranked or more than one spot lower than the misinformation), showing the same findings as reported above.} 
We believe, however, that if there is such impact it would have been exposed in the ``just below'' position, as directly tested in our experiments.
At the same time, our misinformation results could have been skewed by our samples' demographics. We expand on that in Section~\ref{sec:limitations} below. 

At the same time, our results show that the warning-banner approaches to combat misinformation may present some challenges. 
Google’s use of these general warnings~\cite{Sullivan2021} may have been an attempt to avoid the backfire effects of item-specific approaches~\cite{Pennycook2020}, but our findings suggest that general warnings could also have consequences in this context. 
Google's warning about results changing quickly had no effect on trustworthiness evaluations in our experiment, but, as noted above, the warning about unreliable sources reduced trust in results with accurate information (but not those containing misinformation). 
This result 
matches findings from recent studies of general misinformation warnings in news evaluation contexts: such general approaches can backfire by casting doubt on genuinely valid information~\cite{VanDerMeer2023, Greene2021, Clayton2019}. 
Meanwhile, neither warning affected the perceived trustworthiness of misinformation results; however, this null result could be due to a floor effect given that participants in our control conditions already distrusted the misinformation results.

\subsection{Limitations}
\label{sec:limitations}
Our findings were robust across multiple conditions that varied the familiarity of search results’ sources, the medical consensus around the query, the search-engine attribution, and exposure to misinformation and warning banners. Robustness was also demonstrated by the replication of previous results on the rank-click association. That said, our studies also have some important limitations.

First, research has shown that susceptibility to misinformation is concentrated among a small group~\cite{Guess2019, Grinberg2019}. Our participants were drawn from an online sample that did not focus on that group. 
The participants we recruited through Prolific Academic tend to be younger, more politically liberal, and more supportive of vaccination than the overall population. 
The non-representativeness of online samples is a common problem in modern-day survey research~\cite{Bago2020}. 
While we believe our results generalize to a large part of the population, we cannot comment directly on the potential impact of such an experiment on the small group of individuals most susceptible to misinformation.  
Future work should take on the (difficult) task of studying the effects of misinformation in online search and the popular mitigation approaches on such groups.  

We note that our experimental setup prevents us from testing the relationship between higher perceived trustworthiness and increased click probability~(H1) in the presumed causal direction (perceived trustworthiness causes clicks).  
While one can argue that participants may give higher trustworthiness ratings to results they clicked in the previous step as a form of rationalization (a la Self Perception Theory~\cite{Bem1972}), we took steps to mitigate this possibility, including not explicitly reminding participants in the rating step whether the result we selected for them to rate was the one they had clicked. Furthermore, it is possible that rationalization or commitment effects could occur with a reversed design, in which an experiment first measures trustworthiness perceptions of every result on a page, then asks people to click one. In such a design, participants could feel more pressure to click on the result for which they had assigned the highest trustworthiness rating.

Another limitation is that with crisis events, the veracity of information about a novel virus is inherently uncertain, which sometimes clouds the distinction between accurate search results and misinformation. For our study, we used blatant examples of false claims about Covid-19, which were both unsupported by peer-reviewed scientific reports and pretested on separate pools of participants to validate the misinformation condition. We have shown that blatant, detectable misinformation fails to undermine confidence in accurate search results that follow, and this finding can be expected to generalize to other information with uncertain veracity. However, further research is needed to see if these effects of misinformation extend to more subtle forms of misinformation---and differences in trustworthiness---that may appear in search results.

Our experiments were focused on health information, and more specifically Covid-19, but it is possible that results for Covid-19 search queries do not generalize to topics other than health. Health-related searches may require greater deference to professional expertise compared to searches about lifestyle topics, sports, or shopping, for example. 
According to Palen et al., ``When considering emergencies, stakes are often quite high, so credibility must be established quickly, although often only partially, before one decides what to do (or not to do) with [the information]''~\cite{palen2011supporting}. 
Indeed, search in important contexts like health and crisis response is where it is most important to understand, like we do here, the
trustworthiness evaluations and the factors that impact them, for example the presence of misinformation. 
Future work should explore whether our findings hold for other types of searches.

Finally, we ran an (online) lab experiment, and the external validity of such experiment is naturally not as robust as a field experiment. 
Nevertheless, we have some evidence that our treatments worked and that participants responded in thoughtful ways to our interventions (e.g. lower rate of misinformation clicks; lower trustworthiness ratings in Experiment~2 vs. Experiment~3 which had more uncertain queries and results).  
Obviously, the external validity of these results can be improved with a field experiment, e.g. by Google/Alphabet or any other search engine company. 
Such companies can provide the best evidence for the questions we raise here, though with all the challenges and caveats associated with corporate research.\footnote{To not undermine anonymity of this review, we can just say that the researchers here are not affiliated with any of these companies, though, for proper disclosure, we \textit{do} have funding from some of these companies, though not for this project.}
Overall, we believe the manipulation was effective, and our results, while could be enhanced with a larger-scale field study, are informative and extend the current literature on search and trust.



\bibliographystyle{ACM-Reference-Format}
\bibliography{main}


\end{document}